\pdfoutput=1
\documentclass[11pt]{article}
\pdftrailerid{}
\usepackage[utf8]{inputenc}
\usepackage[T1]{fontenc}
\usepackage{lmodern}
\usepackage[margin=1in]{geometry}
\usepackage{microtype}
\usepackage{amsmath,amssymb,amsfonts,array}
\usepackage{booktabs}
\usepackage{graphicx}
\usepackage[font=small,labelfont=bf]{caption}
\usepackage{siunitx}
\usepackage{csquotes}
\usepackage{longtable}
\usepackage{adjustbox}
\usepackage{xcolor}
\usepackage[english]{babel}
\usepackage{algpseudocode}
\usepackage{algorithm}
\usepackage[algo2e]{algorithm2e}
\usepackage[normalem]{ulem}
\usepackage{fancyvrb}
\usepackage[section]{placeins}
\usepackage{authblk}
\usepackage[round]{natbib}
\PassOptionsToPackage{hyphens}{url}
\usepackage{hyperref}
\usepackage{cleveref}
\hypersetup{colorlinks=true, linkcolor=blue!50!black, citecolor=blue!50!black,
  urlcolor=blue!50!black, pdftitle={}, pdfauthor={}, pdfsubject={},
  pdfkeywords={}, pdfcreator={}, pdfproducer={}}

\DefineVerbatimEnvironment{example}{Verbatim}{fontsize=\small}
\newcommand{\sg}[1]{#1}

\title{WMAP: An R Package for Causal Meta-Analysis by Integrating Multiple Observational Studies}
\author[1]{Subharup Guha\thanks{\texttt{subha.guha@dartmouth.edu}}}
\author[2]{Mengqi Xu\thanks{\texttt{m332xu@uwaterloo.ca}}}
\author[1]{Chayce Reed\thanks{\texttt{Chayce.Reed.HSE@dartmouth.edu}}}
\author[3]{Kashish Priyam\thanks{\texttt{kpriyam@ucla.edu}}}
\author[4]{Yi Li\thanks{\texttt{yili@umich.edu}}}
\affil[1]{Department of Biomedical Data Science, Dartmouth College, Hanover, New Hampshire, USA}
\affil[2]{Department of Statistics and Actuarial Science, University of Waterloo, Waterloo, Ontario, Canada}
\affil[3]{Department of Biostatistics, University of California, Los Angeles, Los Angeles, California, USA}
\affil[4]{Department of Biostatistics, University of Michigan, Ann Arbor, Michigan, USA}
\date{}

\begin{document}
\maketitle

\begin{abstract}
Integrating multiple observational studies for meta-analysis has sparked much interest. The presented {R} package WMAP (Weighted Meta-Analysis with Pseudo-Population\sg{s}) \citep{guha2026WMAP} addresses a critical gap in the implementation of integrative weighting approaches for multiple observational studies and causal inferences about various groups of subjects, such as disease subtypes. The package features three weighting approaches, each representing a special case of the unified weighting framework introduced by \cite{guha2024causal}, which includes an extension of inverse probability weights for data integration settings. It performs meta-analysis on user-inputted datasets as follows: (i)  it first estimates the propensity scores for study-group combinations, calculates subject balancing weights, and determines the effective sample size (ESS) for a user-specified weighting method; and (ii)  it then estimates various features of multiple counterfactual group outcomes, such as group medians and differences in group means for the mRNA expression of eight  genes.  Additionally, bootstrap variability estimates are provided. Among the implemented weighting methods, we highlight the \underline{FLEX}ible, \underline{O}ptimized, and \underline{R}ealistic (FLEXOR) method,  which is specifically designed to maximize the ESS within the unified framework. The use of the software is illustrated through simulations and a multi-site breast cancer case study based on a simulated dataset modeled after real TCGA data from seven medical centers.
\end{abstract}

\section{Introduction}
\label{sec:introduction}
When analyzing observational studies, balancing covariates  is a crucial    step for  unconfounded causal comparisons of group potential outcomes  \citep{robins1995semiparametric, Rubin_2007}. In diverse research areas, the \textit{observed} or underlying sampling   populations of observational studies, in addition to being unbalanced with respect to the group-specific covariates, are invariably very different from the larger natural population of interest. It is therefore necessary to utilize covariate-balancing  techniques like weighting or matching  \citep{Lunceford_Davidian_2004}. Weighting methods in observational studies with two subject groups rely on the  propensity score, and overwhelmingly, utilize inverse probability weights (IPWs) to achieve covariate balance \citep{Rosenbaum_Rubin_1983,li2019propensity}. However, IPW estimators of group differences are often  unstable if one or more subjects have extreme PS values \citep{li2019propensity}.   Consequently, variations of IPWs motivated by truncated subpopulations have been developed \citep[e.g.,][]{crump2006moving,li2013weighting}. 

Most weighting methods in the literature  provide valid inferences for a covariate-balanced \textit{pseudo-population}  that differs from the larger, natural population of interest for which no random samples are available. For example, IPWs correspond to a so-called \textit{combined} pseudo-population. Overlap weights minimize the asymptotic variance of the weighted average treatment effect for
the overlap pseudo-population \citep{Li_etal_2018}.    For single observational studies with two or more groups, generalized overlap weights  minimize the  sum of asymptotic variances of   weighted estimators of  
 pairwise group differences \citep{li2019propensity}. 
These approaches have  the following shortcomings: (i)
they are  optimal for  restricted outcome types and estimands (typically, contrasts of group means) under theoretical  conditions that may  not be satisfied in practice; furthermore,
 the study goals may involve very different  estimands than the group mean contrasts of counterfactual outcomes (e.g., percentiles, medians, or     pairwise correlations of multivariate group responses) and various unplanned estimands; (ii) they do not explicitly account for inferential efficiency when multiple admissible pseudo-populations are available, including those constrained to match known characteristics of the larger natural population such as group prevalences; and  (iii) they are not appropriate for the meta-analysis  of multiple observational studies with each study comprising more than two subject groups.

Motivated by these challenges, \cite{guha2024causal} extended  the propensity score  to  the multiple propensity score (MPS). Further, they  proposed a general family of  pseudo-populations and balancing  weights {that facilitate the integrative analyses and 
   causal inferences of diverse  functionals of  group-specific potential outcomes. This unified framework generalizes the aforementioned weighting methods to the meta-analysis of multiple studies with multiple groups. Specifically, as explained in Section \hyperref[sec:methodology]{2}, IPWs and overlap weights \citep{Li_etal_2018, li2019propensity} are extended to  the \textit{integrative combined} (IC) 
and \textit{integrative generalized overlap} (IGO) weights, respectively. Furthermore, by} optimizing  the effective sample size (ESS) of  pseudo-populations within this rich family, \cite{guha2024causal}  invented the so-called 
 \underline{FLEX}ible, \underline{O}ptimized, and \underline{R}ealistic (FLEXOR)
   weights and studied the properties of these estimand-agnostic and  efficient weighted estimators for quantitative, categorical, or multivariate outcomes.  The substantial benefits of FLEXOR relative to extensions of existing weighting methods such as IC and IGO are demonstrated using simulated and TCGA cancer datasets in that paper. The  approach formulated by \cite{guha2024causal} fills a  methodological gap by pioneering a principled, estimand-agnostic integrative causal approach capable of accommodating multiple studies with multiple groups,
   a structure frequently encountered in contemporary integrative research. We have developed an R package, WMAP \citep{guha2026WMAP}, to  implement  the three weighting methods.
      To our knowledge, the present {R} package is the first implementation of  such a method. 
   
   The paper is structured as follows. Section 2  reviews the weighting methodology and two-stage inference procedure of the FLEXOR approach.   Section 3 demonstrates the use of the WMAP package by analyzing the included \texttt{demo} dataset, which mimics the structure of a multi-site breast cancer study from The Cancer Genome Atlas (TCGA). It presents the meta-analysis of retrospective cohorts as a case study, describes the general workflow, illustrates the inferential procedure, and highlights biologically meaningful results. Section 4 conducts simulation studies using the package. 
   Section 5 discusses future directions.

\section{Methodology}
\label{sec:methodology}

In a large population of interest, suppose there are
  $K$ groups  of patients about whom nothing  is known a priori besides the disease  prevalences available from  registries. 
The  meta-analysis integrates $J$  retrospective cohorts or studies  with  each study recording the    covariates for each subject.    For participant $i$, let $S_i$  denote their 
  observational study,  $Z_i$ denote their group, and $\mathbf{X}_i \in \mathcal{X} \subset \mathcal{R}^p$ denote their covariate vector, and vector of $L$ outcomes by $\mathbf{Y}_i\in \mathcal{R}^L$. Writing  the $z$th group's counterfactual 
outcome vector (i.e., the  outcome if the $i$th subject had belonged to group $z$) by 
$\mathbf{Y}_i^{(z)}=(Y_{i1}^{(z)},\ldots,Y_{iL}^{(z)})'$, the observed outcome  is then $\mathbf{Y}_i= \mathbf{Y}_i^{(Z_i)}$.  We further assume that  (a)  $J$ and $K$ are not large, (b) a subject  belongs to just one observational study, and (c) subjects belonging to  all $K$ groups are observed in every study. 

If the subject labels are not meaningful, then $(S_i,Z_i,\mathbf{X}_i,\mathbf{Y}_i)$ may be regarded as  i.i.d.\ samples from an   \textit{observed   population} with density  $p_{+}[S,Z,\mathbf{X},\mathbf{Y}]$, where  $p_{+}[\cdot]$  represents  observed population distributions or densities with respect a dominating measure. 
 Generalizing the assumptions of  \cite{Rubin_2007} and \cite{Imbens_2000}, we assume: (I)~\textit{Stable unit treatment value assumption (SUTVA)}:  Conditional on the covariates, the study and group to which a subject belongs has no effect on their potential outcomes, and every version of grouping would lead to the same potential outcomes; (II) \textit{Study-specific  unconfoundedness}: Given  study $S_i$ and covariate   $\mathbf{X}_i$, group  $Z_i$  is statistically independent of   $\mathbf{Y}_i^{(1)}, \ldots, \mathbf{Y}_i^{(K)}$, i.e., $p_{+}[\mathbf{Y}^{(z)} \mid S,Z, \mathbf{X}] = p_{+}[\mathbf{Y}^{(z)} \mid S, \mathbf{X}]$; and (III) \textit{Positivity}: $p_{+}[S=z,Z=z,\mathbf{X}=\mathbf{x}]>0$  for all $(s,z,\mathbf{x})$.
  Under these conditions, the  presented WMAP   package performs a two-stage analysis: Stage 1 computes weights to integrate multiple studies, and Stage 2 uses these weights to infer group counterfactual outcomes.

\subsection{Stage 1: Outcome-free sample weights}\label{S:stage1}
For the purpose of meta-analysis, we  generalize  the propensity score \citep[e.g.,][]{Rosenbaum_Rubin_1983}   to  the  {\textit{multiple propensity score} (MPS)} of 
  study-group memberships belonging to $\Sigma=$
 $\{1,\ldots,J\} \times \{1,\ldots,K\}.$ For  $\mathbf{x} \in \mathcal{X}$, the MPS is defined as
\begin{equation}
    \delta_{sz}(\mathbf{x})    =  
p_{+}\bigl[S=s,Z=z \mid \mathbf{X} = \mathbf{x}\bigr]\quad\text{for   $(s,z) \in \Sigma$.
} \label{o-MPS}
\end{equation}
In observational studies, the unknown MPS is estimated by regressing the  factor variable $(S_i,Z_i)$ on $\mathbf{x}_i$   for the $N$ subjects using  parametric or nonparametric regression techniques. Using MPS, we can derive subject-specific balancing weights that redistribute, and thereby, transform the density of the (unbalanced) observed population  to 
a covariate-balanced \textit{pseudo-population} \citep{guha2024causal} in which a patient's
 study,  group  and   covariates are mutually independent by design: 
\begin{equation}
   p\bigl[S=s, Z=z,\mathbf{X}=\mathbf{x}\bigr]= \gamma_s \, \theta_z \,  f_{\boldsymbol{\gamma},\boldsymbol{\theta}}(\mathbf{x}),  \quad \text{for $(s,z,\mathbf{x}) \in \Sigma\times\mathcal{X}$,}\label{pseudo1}
\end{equation}
 where $p[\cdot]$ denotes distributions or  densities  with respect to the pseudo-population,  and which relies on probability vector $\boldsymbol{\gamma}=(\gamma_1,\ldots,\gamma_J)$ quantifying the study relative masses, relative group prevalences $\boldsymbol{\theta}=(\theta_{1},\ldots,\theta_{K})$, and  pseudo-population covariate density $f_{\boldsymbol{\gamma},\boldsymbol{\theta}}(\mathbf{x})$. In some investigations, it is possible to specify  group prevalence $\boldsymbol{\theta}$ to match the group prevalence of the natural population. However, reasonable choices of $\boldsymbol{\gamma}$ are often unknown because they are primarily determined by the study designs and unknown factors influencing cohort participation. If  any component of $\boldsymbol{\gamma}$ or $\boldsymbol{\theta}$ is unknown, we  can optimize the pseudo-population in later steps with respect to these quantities.

Let the  marginal observed covariate density be denoted by $f_{+}(\mathbf{x})$,  $\mathbf{x}\in \mathcal{X}$.
 Then,  there exists a  \textit{tilting function} \citep[e.g.,][]{Li_etal_2018}, $\eta_{\boldsymbol{\gamma},\boldsymbol{\theta}}$,    such that
$f_{\boldsymbol{\gamma},\boldsymbol{\theta}}(\mathbf{x}) \propto \eta_{\boldsymbol{\gamma},\boldsymbol{\theta}}(\mathbf{x})f_{+}(\mathbf{x})$, implying that $f_{\boldsymbol{\gamma},\boldsymbol{\theta}}(\mathbf{x}) = \eta_{\boldsymbol{\gamma},\boldsymbol{\theta}}(\mathbf{x})f_{+}(\mathbf{x})/\mathbb{E}_+[\eta_{\boldsymbol{\gamma},\boldsymbol{\theta}}(\mathbf{X})]$, where   $\mathbb{E}_+(\cdot)$ denotes expectations under the observed distribution in which $\mathbf{X} \sim f_{+}$.  This gives an alternative characterization of pseudo-populations relying on tilting functions instead of covariate densities. In particular, with $\mathcal{S}_{J}$ denoting the unit simplex in $\mathcal{R}^J$, different    $\boldsymbol{\gamma}\in\mathcal{S}_{J}$, $\boldsymbol{\theta}\in\mathcal{S}_{K}$,  and  tilting function $\eta_{\boldsymbol{\gamma},\boldsymbol{\theta}}$     result in different  pseudo-populations belonging to  family (\ref{pseudo1}) as natural meta-analytical extensions of  
existing   weighting methods  for single studies.  For example,   equally weighted studies and  groups, along with tilting function $\eta_{\boldsymbol{\gamma},\boldsymbol{\theta}}(\mathbf{x})\propto 1$ and  $\eta_{\boldsymbol{\gamma},\boldsymbol{\theta}}(\mathbf{x}) =1/\sum_s\sum_z\delta_{sz}^{-1}(\mathbf{x})$, respectively,  extends the combined \citep{Li_etal_2018} and generalized overlap \citep{li2019propensity} pseudo-populations  to  the integrative combined (IC) 
and integrative generalized overlap (IGO) pseudo-populations, respectively.

For more generality and efficiency, we  define a   \textit{(multi-study) balancing  weight} 
as the ratio of   pseudo-population and  observed population densities \citep{guha2024causal}.  For  $(s,z,\mathbf{x}) \in \Sigma\times\mathcal{X}$,  balancing  weight 
\begin{equation}
    \rho_{\boldsymbol{\gamma},\boldsymbol{\theta}}(s,z,\mathbf{x}) = %
\frac{\gamma_s \, \theta_z  \, \eta_{\boldsymbol{\gamma},\boldsymbol{\theta}}(\mathbf{x})}{\delta_{sz}(\mathbf{x}) \,\mathbb{E}_+[\eta_{\boldsymbol{\gamma},\boldsymbol{\theta}}(\mathbf{X})]}. 
\label{w1}
\end{equation}
 redistributes the   observed distribution's density to equal   the pseudo-population's density.  The \textit{unnormalized sample weights}, $\tilde{\rho}_{i}=$ $\gamma_{s_i}  \theta_{z_i}   \eta_{\boldsymbol{\gamma},\boldsymbol{\theta}}(\mathbf{x}_i)/\delta_{s_iz_i}(\mathbf{x}_i)$,  are later used to evaluate  unconfounded  weighted estimators of various  counterfactual outcome features in the pseudo-population. The \textit{empirically normalized balancing weight} is computed from the $N$ unnormalized sample weights as $\rho_i=\tilde{\rho}_{i}/\sum_{u=1}^N \tilde{\rho}_{u}$ and is normalized to have  sample mean  1. 

A widely used, estimand-agnostic measure of a weighting method's inferential accuracy is the \textit{effective sample size} (ESS) \citep[e.g.,][]{mccaffrey2013tutorial}, 
$  \mathcal{Q}(\boldsymbol{\gamma}, \boldsymbol{\theta},\eta_{\boldsymbol{\gamma},\boldsymbol{\theta}}) = N/\bigl[1+\text{Var}_{+}\bigl\{\rho_{\boldsymbol{\gamma},\boldsymbol{\theta}}(S,Z,\mathbf{X})\bigr\}\bigr] 
    = N/\mathbb{E}_{+}\bigl\{\rho_{\boldsymbol{\gamma},\boldsymbol{\theta}}^2(S,Z,\mathbf{X})\bigr\}$, where  $\mathbb{E}[\cdot]$ and $\text{Var}[\cdot]$ respectively  denote expectations and variances with respect to the  pseudo-population. When  $N$ is large, the ESS is reliably estimated by the sample ESS,  $\tilde{\mathcal{Q}}(\boldsymbol{\gamma}, \boldsymbol{\theta},\eta_{\boldsymbol{\gamma},\boldsymbol{\theta}}) = N^2/\sum_{i=1}^N \rho_{i}^2$, using the empirically normalized weights.

We define the FLEXOR pseudo-population as a member of family (\ref{pseudo1}) maximizing ESS subject to any investigator-imposed restrictions that $(\boldsymbol{\gamma},\boldsymbol{\theta}) \in \Upsilon \subset \mathcal{S}_J \times \mathcal{S}_K$ \citep{guha2024causal}. The constraint set $\Upsilon$ can be chosen to preserve known characteristics of the target population, such as group prevalences, while excluding implausible study-mass allocations. If  the FLEXOR pseudo-population is identified by  $(\breve{\boldsymbol{\gamma}}, \breve{\boldsymbol{\theta}},\breve{\eta}_{\breve{\boldsymbol{\gamma}}, \breve{\boldsymbol{\theta}}})$, then 
\[\mathcal{Q}(\breve{\boldsymbol{\gamma}}, \breve{\boldsymbol{\theta}},\breve{\eta}_{\breve{\boldsymbol{\gamma}}, \breve{\boldsymbol{\theta}}}) = \sup_{(\boldsymbol{\gamma}, \boldsymbol{\theta}) \in \Upsilon} \sup_{\eta_{\boldsymbol{\gamma},\boldsymbol{\theta}}} \mathcal{Q}(\boldsymbol{\gamma}, \boldsymbol{\theta},\eta_{\boldsymbol{\gamma},\boldsymbol{\theta}}).
\]

\paragraph{An iterative procedure for estimating FLEXOR pseudo-population}  
 Initializing  $(\boldsymbol{\gamma}, \boldsymbol{\theta})\in\Upsilon$,  we  
 perform the following   steps iteratively until the sample ESS converges. The converged pseudo-population  with optimized sample ESS  estimates the FLEXOR pseudo-population.

\begin{itemize}
    \item   With  the parameters $(\boldsymbol{\gamma} , \boldsymbol{\theta})$ held fixed, maximize the sample ESS $\tilde{\mathcal{Q}}(\boldsymbol{\gamma}, \boldsymbol{\theta},\eta_{\boldsymbol{\gamma},\boldsymbol{\theta}})$ over all  tilting functions to obtain the \textit{best fixed-$(\boldsymbol{\gamma}, \boldsymbol{\theta})$ pseudo-population} represented by   $(\boldsymbol{\gamma}, \boldsymbol{\theta},\breve{\eta}_{\boldsymbol{\gamma},\boldsymbol{\theta}})$.  The analytical form of $\breve{\eta}_{\boldsymbol{\gamma},\boldsymbol{\theta}}$  {for the theoretical ESS} is given by Theorem 1 of \cite{guha2024causal}: 
    \[       \breve{\eta}_{\boldsymbol{\gamma},\boldsymbol{\theta}}(\mathbf{x}) =   \biggl( \sum_{s=1}^J  \sum_{z=1}^K \frac{\gamma_{s} ^2\theta_z^2}{\delta_{sz}(\mathbf{x})}  \biggr)^{-1},  \quad \text{ $\mathbf{x} \in \mathcal{X}$.}
  \]
   This pseudo-population's  balancing  weights (\ref{w1})     are uniformly bounded over $(s,z,\mathbf{x}) \in \Sigma\times\mathcal{X}$. Tilting function $\breve{\eta}_{\boldsymbol{\gamma},\boldsymbol{\theta}}(\mathbf{x})$ de-emphasizes  covariate regions where $\delta_{sz}(\mathbf{x})$ is nearly 0 for some $(s,z) \in \Sigma$. At the same time, it  promotes  covariate regions where the  group propensities, 
 $\delta_{z}=\sum_{s=1}^K \delta_{sz}(\mathbf{x})$, match  group proportion $\theta_z$
for every $z=1,\ldots,K$. %
    Set function $\eta=\breve{\eta}_{\boldsymbol{\gamma},\boldsymbol{\theta}}$.
  
  \item    With  tilting function $\eta$ held fixed, maximize  the sample ESS {$\tilde{\mathcal{Q}}(\boldsymbol{\gamma}, \boldsymbol{\theta},\eta)$} over all parameters $(\boldsymbol{\gamma}, \boldsymbol{\theta})\in\Upsilon$ to obtain  the  \textit{best fixed-$\eta$ pseudo-population} represented by $(\tilde{\boldsymbol{\gamma}}, \tilde{\boldsymbol{\theta}},\eta)$. The   numerical maximization can be  performed  using implementations of   Gauss-Seidel or  Jacobi  algorithms.  Set $(\boldsymbol{\gamma}, \boldsymbol{\theta})=(\tilde{\boldsymbol{\gamma}}, \tilde{\boldsymbol{\theta}})$.
 
\end{itemize}

 The WMAP package includes   a function, \texttt{balancing.weights()}, that estimates the MPS and calculates the normalized balancing weights of the $N$ subjects and  sample ESS for a user-specified pseudo-population. The required pseudo-population is specified by the user through the argument {\tt method}, which can  be ``FLEXOR'', ``IC'', or ``IGO.'' If {\tt method="FLEXOR"}, the  iterative procedure, as outlined above, is implemented starting from different initial values to estimate the FLEXOR pseudo-population.

\subsection{Stage 2: Unconfounded inferences of  group counterfactual outcomes}\label{S:stage2}

The theoretical properties (e.g., asymptotic variances) of multivariate weighted estimators of wide-ranging group-level features of the outcomes have been thoroughly investigated  \citep{guha2024causal}. We summarize here some special applications of the methodology relevant to the WMAP package implementation.
 Using univariate outcomes ($L=1$),  suppose   the 
  potential outcome vectors $Y^{(1)},\ldots,$ $Y^{(K)}$ have  common support, $\mathcal{Y}\subset\mathcal{R}$.  We  assume  identical conditional distributions: $p[Y^{(z)} \mid S,Z, \mathbf{X}] = p_{+}[Y^{(z)} \mid S,Z,  \mathbf{X}]$  for  $z=1,\ldots,K$,  guaranteeing   that the SUTVA,  unconfoundedness, and positivity assumptions  for the observed population    hold for  the pseudo-population.  Since the   pseudo-population has balanced covariates by design, this implies that $
   p [Y \mid Z=z] = p[Y^{(z)}]$,
paving the way for weighted estimators of   potential outcome features for the FLEXOR, IGO, and IC pseudo-populations. 

 Let $\Phi_1,\ldots,\Phi_M$  be real-valued 
 functions having domain $\mathcal{Y}$. We wish to infer 
 pseudo-population means of   transformed potential  outcomes, $\mathbb{E}[\Phi_1(Y^{(z)})],\ldots,\mathbb{E}[\Phi_M(Y^{(z)})]$  for   $z=1,\ldots, K$.  Appropriate choices of   $\Phi_m$ correspond to pseudo-population inferences about group-specific  marginal
means, medians,  variances, and  CDFs of     potential outcome components. 
Equivalently, writing $\mathbf{\Phi}(Y^{(z)})=\bigl(\Phi_1(Y^{(z)}),\ldots,\Phi_M(Y^{(z)})\bigr)' \in \mathcal{R}^M$, let $\boldsymbol{\lambda} ^{(z)}=$ $\mathbb{E}[\mathbf{\Phi}(Y^{(z)})]$. 
 For  real-valued functions $\psi$ with domain $\mathcal{R}^M$, 
we wish to estimate $\psi(\boldsymbol{\lambda} ^{(z)})$. 

For example,  define  $\Phi_1(Y^{(z)})=Y^{(z)}$ and  $\Phi_2(Y^{(z)})=(Y^{(z)})^2$. Then    $\psi(t_1,t_2)=t_1$, we obtain the pseudo-population mean in the  $z$th group. Similarly, $\psi(t_1,t_2)=\sqrt{t_2-t_1^2}$ gives the pseudo-population standard deviation in the  $z$th group. For a second example, let $y_{1},\ldots,y_{M}$ be  a fine grid of prespecified points in the support $\mathcal{Y}$ and  $\Phi_m(Y^{(z)})=\mathcal{I}(Y^{(z)}\le y_{m})$. For  $\psi(t_1,\ldots,t_M)=t_{m}$,   the pseudo-population CDF of $Y^{(z)}$ evaluated at $y_{m}$ equals  $\psi(\boldsymbol{\lambda} ^{(z)})$. Similarly, for  $\psi(t_1,\ldots,t_M)=t_{m^*}$ where $m^*=\arg\min_m |t_m-0.5| $,  the approximate median of $Y_1^{(z)}$ is given by $\psi(\boldsymbol{\lambda} ^{(z)})$. 
 
Using the normalized   weights $\rho_1,\ldots, \rho_N$,  
a weighted estimator of mean  vector $\boldsymbol{\lambda} ^{(z)}$ is
\begin{equation}
    \mathbf{\bar{\Phi}}_{z} = \frac{\sum_{i=1}^N \rho_i\,\mathbf{\Phi}(Y_i)\,\mathcal{I}(Z_i=z)}{\sum_{i=1}^N \rho_i\,\mathcal{I}(Z_i=z)}, \label{estimator}
\end{equation}
and $\psi(\mathbf{\bar{\Phi}}_{z})$ is an  estimator of $\psi(\boldsymbol{\lambda}^{(z)})$. Theorem 2 of \cite{guha2024causal}  shows that these weighted estimators are consistent and asymptotically normal. However, since $N$ may not be sufficiently large to justify asymptotic approximations, the WMAP package applies bootstrap methods to estimate  standard errors.  

As an optional outcome-modeling extension, WMAP allows users to replace observed outcomes with predictions from random forest models fit using covariates, study membership, and group membership. The weighted estimators are then computed using these predicted outcomes in place of the observed responses. This approach can improve precision when outcomes are strongly associated with the observed predictors. To reduce overfitting, cross-fitted predictions may also be used.

The Stage 2 analysis is implemented in WMAP by function \texttt{causal.estimate()}, which first calls function \texttt{balancing.weights()}  to perform the Stage 1 analyses.
The function then  estimates different features of the $K$ counterfactual  group outcomes (e.g., medians and group mean differences) for the IC, IGO, or FLEXOR pseudo-populations. The  function also evaluates  bootstrap-based variability estimates for these  features. 

FLEXOR differs from IC and IGO in that it explicitly maximizes the ESS over a class of admissible pseudo-populations, including those matching known group prevalences in the natural population and restricting the study-mass parameters to scientifically plausible ranges. Because larger ESS generally corresponds to more stable weights and lower estimator variability, FLEXOR often yields more precise inferences and narrower confidence intervals in finite samples. The approach is particularly attractive when the scientific goals involve a broad range of estimands rather than a single target estimand for which alternative weighting methods may be theoretically optimal. As with all weighting-based methods, however, FLEXOR remains sensitive to limited covariate overlap and small study-group sample sizes, situations in which propensity score estimation and weighted inference can become unstable.

More broadly, the proposed integrative weighting framework is particularly useful when investigators seek to integrate evidence across multiple observational studies and make inference for a target population defined jointly by study membership and group assignment. In such settings, weighting methods based solely on treatment propensity scores may not adequately account for differences in study composition. By constructing study- and group-balanced pseudo-populations, the framework enables causal inference for scientifically meaningful target populations while optimizing inferential efficiency within a broad class of admissible pseudo-populations.

\section{Meta-analysis of multiple (multi-site) breast cancer studies}
\label{sec:datasets} 

To demonstrate the use of the WMAP package, we analyze the built-in \texttt{demo} dataset, which contains simulated data designed to mirror the structure and characteristics of a multi-site breast cancer study from The Cancer Genome Atlas (TCGA). This illustrative dataset mimics the layout and variable types of the original TCGA study, which involved $J = 7$ medical centers and $N = 450$ patients, divided into $K = 2$ groups based on breast cancer subtypes: infiltrating ductal carcinoma (IDC) and infiltrating lobular carcinoma (ILC). The original dataset, available from the GDC Data Portal upon registration \citep{GDC}, includes $p = 30$ unbalanced covariates and $L = 8$ outcomes representing mRNA expression levels for selected genes (COL9A3, CXCL12, IGF1, ITGA11, IVL, LEF1, PRB2, and SMR3B) known to be relevant in breast cancer research \citep{christopoulos2015role}. Using the \texttt{demo} dataset, we estimate and compare the counterfactual means, standard deviations, and medians between the IDC and ILC groups. The WMAP package thus provides a convenient and accessible tool for exploring causal meta-analysis methods on synthetic datasets that closely resemble real-world cancer studies, even without direct access to the full TCGA resource.

\subsection{Data structure} 
We begin by  installing and loading the WMAP package and the example dataset included in the package.
\begin{example}
R> library(WMAP)
R> data(demo)
\end{example}

We then examine the contents of the dataset.
\begin{itemize}
    \item \texttt{X}: Continuous/binary covariates --- $p=30$ demographic and clinicopathological covariates.
    \begin{example}
R> dim(X)
[1] 450  30
        \end{example}
    \item \texttt{Y}: Continuous outcome --- outcome vectors of mRNA expression measurements for the eight targeted genes arranged in a $450 \times 8$ matrix.
    \begin{example}
R> round(head(Y),4)
        [,1]    [,2]    [,3]    [,4]    [,5]    [,6]    [,7]    [,8]
[1,]  1.2828 -0.1152 -0.3829 -0.3082 -1.1200  1.2068 -0.9472  0.8768
[2,] -1.1603  1.5377  1.6034 -0.9822 -0.9507  0.3980 -0.9481 -0.2325
[3,] -0.3815  1.1320  0.9348 -1.2661  1.1733 -0.0956 -0.1138  2.3797
[4,] -0.3032  0.5999  1.3941 -0.0299 -1.1010 -0.0838 -0.9565 -0.5058
[5,]  0.4147 -0.3312 -1.7675  0.7626 -1.1300 -0.6157 -0.4613 -0.9950
[6,]  0.1867  0.9826  1.2274  0.7895  0.2217 -0.5541 -0.9479 -1.0016
        \end{example}
    \item \texttt{S}: Categorical vectors --- site labels of patients, representing the seven medical centers.
    \item \texttt{Z}: Categorical vectors --- group labels of patients, representing the two disease subtypes IDC and ILC. 
    \item \texttt{groupnames:}
    \begin{example}
R> groupNames
[1] "Infiltrating Ductal Carcinoma" 
[2] "Infiltrating Lobular Carcinoma"
        \end{example}
    \item \texttt{naturalGroupProp:} The relative proportions of IDC and ILC subtypes in the larger U.S. population  \citep{IDC,ILC}.
    \begin{example}
R> naturalGroupProp
[1] 0.8888889 0.1111111
        \end{example}
\end{itemize}

\noindent {\bf Remark:} Users can easily utilize the package functions to conduct meta-analyses on their own datasets.   The formatting requirements for user-specified datasets are as follows:  (a) Vector {\tt S}, consisting of $N$ factor levels belonging to the set $\{1,\ldots,J\}$, representing the  study memberships of the subjects. Each study must have at least 1 subject; (b)   Vector {\tt Z}, consisting of  factor levels belonging to the set $\{1,\ldots,K\}$, representing the $N$ group memberships. Each group should  contain at least 1 subject; (c)  Covariate matrix, {\tt X}, of dimension $N \times p$ containing $p$ continuous or binary (0/1) measurements, including  factor covariates  expanded as dummy binary values;   (d)   Matrix {\tt Y} of dimension $N\times L$ comprising $L$ containing outcomes; and (e) Probability vector, {\tt naturalGroupProp}, of length $K$ and strictly positive elements, representing the relative group prevalences $\boldsymbol{\theta}$ of the larger natural population. This last user input is necessary only for FLEXOR weights. It should be skipped for IC and IGO weights, which assume $\boldsymbol{\theta}=(1/K,\ldots,1/K)$; if specified, the input is ignored  for these weighting methods. 

\begin{algorithm}[!h]
\caption{Stage 1 analysis: Empirically normalized balancing weights}  
\label{alg:stage1}

\SetAlgoLined
\DontPrintSemicolon
\textbf{Input$_1$:} {{\tt S}, {\tt Z},   {\tt X},  {\tt method},  {\tt seed},  {\tt naturalGroupProp},  {\tt num.random},  {\tt gammaMin},  {\tt gammaMax}  

\Comment{See Table \ref{tab:stage1 inputs} for input argument details}}    

    \SetKwFunction{stageone}{balancing.weights}
    \SetKwProg{Fn}{Function}{:}{}
    \Fn{\stageone{\textbf{Input$_1$}}}{

    Regress     {\tt (S,Z)} on   {\tt X} and evaluate  $\hat{\delta}_{s_1z_1}(\mathbf{x}_1),\ldots,\hat{\delta}_{s_Nz_N}(\mathbf{x}_N)$ \Comment{Estimate MPS}

    \If{{\tt method="IC"}} 
    {
    \For{$i=1,\ldots,N$}{
     $\tilde{\rho}_{i}\gets$ $1/\hat{\delta}_{s_iz_i}(\mathbf{x}_i)$ \Comment{IC unnormalized  weights}
    }
    }

    \If{{\tt method="IGO"}} 
    {
    \For{$i=1,\ldots,N$}{
    $\eta(\mathbf{x}_i) \gets 1/\sum_{s=1}^J\sum_{z=1}^K\hat{\delta}_{sz}^{-1}(\mathbf{x}_i)$
     \Comment{IGO tilting function}

     $\tilde{\rho}_{i}\gets$ $\eta(\mathbf{x}_i)/\hat{\delta}_{s_iz_i}(\mathbf{x}_i)$ \Comment{IGO unnormalized  weights}
    }
    }

    \If{{\tt method="FLEXOR"}} 
    {
    $\boldsymbol{\theta} \gets$ {\tt naturalGroupProp}\Comment{fixed group prevalence} 
    
    \For{$t=1,\ldots,\text{\tt num.random}$} 
    {
    
    $\boldsymbol{\gamma}^{(t)} \stackrel{i.i.d.}\sim \mathcal{S}_J \cap \bigl([\text{\tt gammaMin}, \text{\tt gammaMax}\bigr)]^J$
    \Comment{random starting point}

    $\boldsymbol{\gamma}^{(t)}_{\dag}, \mathcal{Q}^{(t)} \gets$ FLEXOR.2STEP$\bigl(\cdots, \boldsymbol{\theta},\{\hat{\delta}_{s_iz_i}(\mathbf{x}_i)\}_{i=1}^N, \boldsymbol{\gamma}^{(t)}\bigr)$ 
    
    \Comment{See Algorithm \ref{alg:FLEXOR2step}}

    } %

    $\breve{t} \gets \text{argmax}_{t}\mathcal{Q}^{(t)}$
    
    $\breve{\boldsymbol{\gamma}} \gets \boldsymbol{\gamma}^{(\breve{t})}$
    
    $\mathcal{Q}\gets$  OPTIMIZED.ESS$\bigl(\breve{\boldsymbol{\gamma}},\boldsymbol{\theta}, \{\hat{\delta}_{s_iz_i}(\mathbf{x}_i)\}_{i=1}^N\bigr)$
    \Comment{See Algorithm \ref{alg:FLEXOR2step}}

    \For{$i=1,\ldots,N$}{
    $\eta(\mathbf{x}_i) \gets$ $\biggl( \sum_{s=1}^J  \sum_{z=1}^K \frac{\breve{\gamma}_{s} ^2\theta_z^2}{\delta_{sz}(\mathbf{x}_i)}  \biggr)^{-1}$ 
    
    \Comment{ Best fixed-$(\boldsymbol{\gamma}, \boldsymbol{\theta})$ tilting function}
    } %

    \For{$i=1,\ldots,N$} {
    $\tilde{\rho}_{i}\gets$ $\breve{\gamma}_{s_i}  \theta_{z_i}   \eta(\mathbf{x}_i)/\hat{\delta}_{s_iz_i}(\mathbf{x}_i)$ \Comment{FLEXOR unnormalized  weights}
    } %
    
    \For{$i=1,\ldots,N$} {
    $\rho_i=\tilde{\rho}_{i}/\sum_{u=1}^N \tilde{\rho}_{u}$ \Comment{Empirically normalized  weights}
    } %
    } %

    \textbf{return} Empirically normalized  weights, {\tt wt.v} $\equiv \{\rho_{i}\}_{i=1}^N$ and  {\tt percentESS} $\equiv 100\mathcal{Q}/N$; see Table \ref{tab:stage1 outputs} for  output details.
}

\textbf{End Function}
\end{algorithm}

\begin{algorithm}[h]
\caption{Iterative procedure of estimating FLEXOR pseudo-population} 
\label{alg:FLEXOR2step}
\SetAlgoLined
\DontPrintSemicolon

\SetKwFunction{FMain}{FLEXOR.2STEP}
\SetKwProg{Fn}{Function}{:}{}
    \KwIn{{\tt S}, {\tt Z},   {\tt X},   {\tt gammaMin},  {\tt gammaMax}, $\boldsymbol{\theta},\{\hat{\delta}_{s_iz_i}(\mathbf{x}_i)\}_{i=1}^N$, $\boldsymbol{\gamma}^{(t)}$ \Comment{passed by Algorithm \ref{alg:stage1}}}  
    \Fn{\FMain{\textbf{Input}}}{

        $\boldsymbol{\gamma} \gets \boldsymbol{\gamma}^{(t)}$; $\mathcal{Q}^{(\text{new})}=0$; {\tt exit=FALSE} \Comment{Initialize}

        $\boldsymbol{\Gamma} \gets \mathcal{S}_J \cap \bigl([\text{\tt gammaMin}, \text{\tt gammaMax}\bigr)]^J$ \Comment{Admissible values of $\boldsymbol{\gamma}$}

        \While{\text{\tt !exit}}{
            $\mathcal{Q}^{(\text{old})}\gets \mathcal{Q}^{(\text{new})}$
            
            $\boldsymbol{\gamma}^{\dag} \gets \text{argmax}_{\boldsymbol{\gamma} \in \boldsymbol{\Gamma}}$ OPTIMIZED.ESS$\bigl(\boldsymbol{\gamma},\boldsymbol{\theta}, \{\hat{\delta}_{s_iz_i}(\mathbf{x}_i)\}_{i=1}^N\bigr)$ \Comment{function defined below}
            
            $\mathcal{Q}^{(\text{new})}=$ OPTIMIZED.ESS$\bigl(\boldsymbol{\gamma}^{\dag},\boldsymbol{\theta}, \{\hat{\delta}_{s_iz_i}(\mathbf{x}_i)\}_{i=1}^N\bigr)$

            \If{$\mathcal{Q}^{(\text{new})}/\mathcal{Q}^{(\text{old})}-1$ is small} {
             {\tt exit} $\gets$  TRUE
            }

} 

        \textbf{return} $\boldsymbol{\gamma}^{\dag}, \mathcal{Q}^{(\text{new})}$
}
\textbf{End Function}

  \vspace{20 pt}

\SetKwFunction{Optimized}{Optimized.ESS}
\SetKwProg{Fn}{Function}{:}{}
    \Fn{\Optimized{$\boldsymbol{\gamma},\boldsymbol{\theta}, \{\hat{\delta}_{s_iz_i}(\mathbf{x}_i)\}_{i=1}^N$}}
    {\Comment{Maximum ESS for pseudo-population parameters  $(\boldsymbol{\gamma},\boldsymbol{\theta})$}

        \For{$i=1,\ldots,N$}{
$\breve{\eta}_{\boldsymbol{\gamma},\boldsymbol{\theta}}(\mathbf{x}_i) \gets$ $\biggl( \sum_{s=1}^J  \sum_{z=1}^K \frac{\gamma_{s} ^2\theta_z^2}{\delta_{sz}(\mathbf{x}_i)}  \biggr)^{-1}$ 

\Comment{ Best fixed-$(\boldsymbol{\gamma}, \boldsymbol{\theta})$ tilting function}
} %

    \textbf{return} Sample.ESS$\bigl(\boldsymbol{\gamma},\boldsymbol{\theta}, \breve{\eta}, \{\hat{\delta}_{s_iz_i}(\mathbf{x}_i)\}_{i=1}^N\bigr)$ \Comment{function defined below}
}      

\textbf{End Function}

  \vspace{20 pt}

\SetKwFunction{Sample}{Sample.ESS}
\SetKwProg{Fn}{Function}{:}{}
    \Fn{\Sample{$\boldsymbol{\gamma},\boldsymbol{\theta}, \eta, \{\hat{\delta}_{s_iz_i}(\mathbf{x}_i)\}_{i=1}^N$}}
    {
        \For{$i=1,\ldots,N$} {
         $\tilde{\rho}_{i}\gets$ $\gamma_{s_i}  \theta_{z_i}   \eta(\mathbf{x}_i)/\hat{\delta}_{s_iz_i}(\mathbf{x}_i)$ \Comment{unnormalized  weights}
        } %

        \For{$i=1,\ldots,N$} { $\rho_i=\tilde{\rho}_{i}/\sum_{u=1}^N \tilde{\rho}_{u}$ \Comment{Empirically normalized  weights}} %

        $\tilde{\mathcal{Q}}(\boldsymbol{\gamma}, \boldsymbol{\theta},\breve{\eta}_{\boldsymbol{\gamma},\boldsymbol{\theta}}) \gets N^2/\sum_{i=1}^N \rho_{i}^2$ 
        
    \textbf{return} $\tilde{\mathcal{Q}}(\boldsymbol{\gamma}, \boldsymbol{\theta},\eta_{\boldsymbol{\gamma},\boldsymbol{\theta}})$
 
 } 

\textbf{End Function}

\end{algorithm}

 \begin{table}[h]
 \setlength{\abovecaptionskip}{2pt}
 \centering
 \renewcommand{\arraystretch}{0.95}
\caption{
 Input arguments for function \texttt{balancing.weights()}.}
 \label{tab:stage1 inputs}
\begin{tabular}{lll}
\\
  \toprule
  \textbf{Argument}
  &\textbf{Short description} &\textbf{Default} 
		\\

\midrule
{\tt S} & Vector of factor levels representing the $N$  study &-\\ &  memberships.  Takes values in $\{1,\ldots,J\}$ \\
{\tt Z} & Vector of factor levels representing the $N$  &-\\
& group memberships.  Takes values in $\{1,\ldots,K\}$ \\
{\tt X} & Covariate matrix of $N$ rows and $p$ columns &-  \\
{\tt method} & Pseudo-population, i.e., weighting method;   &-  \\
&Can  be {\tt "FLEXOR"}, {\tt "IC"}, or {\tt "IGO"} \\
{\tt seed} & Seed for random number generation   &{\tt NULL}  \\
\midrule
\multicolumn{3}{c}{Relevant only when {\tt method="FLEXOR"}; inputs  ignored otherwise}\\
\midrule
{\tt naturalGroupProp} & Relevant only for FLEXOR pseudo-populations: &-\\
&fixed user-specific probability vector $\boldsymbol{\theta}$ \\
{\tt num.random} & Number of random starting points of $\boldsymbol{\gamma}$ &40\\
& in the  iterative procedure\\
{\tt gammaMin} &Lower bound for each $\gamma_s$ in the iterative procedure &0.001\\
{\tt gammaMax} &Upper bound for each $\gamma_s$ in the iterative procedure &0.999\\
	 \bottomrule
\end{tabular}
\end{table}

\begin{table}[h]
 \setlength{\abovecaptionskip}{2pt}
 \centering
\caption{
 Output list components of   function \texttt{balancing.weights()}.}
 \label{tab:stage1 outputs}
\begin{tabular}{lll}
\\
  \toprule
  \textbf{Position}&\textbf{Names} 
  &\textbf{Short description} 
		\\

\midrule
1 & {\tt wt.v} &$N$ empirically normalized sample weights\\
2 & {\tt percentESS} &Percentage sample ESS for pseudo-population\\
3 & {\tt method} & Weighting method used (IC, IGO, or FLEXOR) \\
4 & {\tt selected\_nodesize} & Selected {\tt min.node.size} for the propensity forest \\
5 & {\tt diagnostics} & List of propensity model diagnostics, containing \\
  &  & $\bullet$ {\tt frac\_truncated}: fraction of subjects with near-zero \\ 
  &  & pre-truncation propensity \\
  &  & $\bullet$ {\tt crossfit\_accuracy}: OOF classification accuracy of \\ 
  &  & propensity model \\
  &  & $\bullet$ {\tt max\_weight}: maximum balancing weight \\
  &  & $\bullet$ {\tt prob\_quantiles}: quantiles of estimated propensity scores \\
	 \bottomrule
\end{tabular}
\end{table}

\subsection{Workflow of analysis}
\label{sec:workflow}

\subsubsection[Stage 1: balancing.weights()]{Stage 1: {\tt balancing.weights()}}
For a prespecified pseudo-population,
 function \texttt{balancing.weights()} first estimates the MPS and then calculates the subject-specific normalized balancing weights and  sample ESS. The input arguments are summarized in Table \ref{tab:stage1 inputs}. The workflow,  outlined in Algorithm \ref{alg:stage1},  is based on the iterative \hyperref[S:stage1]{Stage 1} procedure described in Section~\hyperref[sec:methodology]{2} and detailed in Algorithm \ref{alg:FLEXOR2step}. The function returns a list of items summarized in Table \ref{tab:stage1 outputs}.
 If {\tt method} is ``IC'' or ``IGO,'' many arguments of \texttt{balancing.weights()} are not required and any user-provided values are ignored. For instance,  $\gamma_s=1/J$ and $\theta_z=1/K$ are fixed for these pseudo-populations.

 For the FLEXOR pseudo-population, (i) the function assumes that the $K$ group proportions $\boldsymbol{\theta}$ are fixed and specified by the user in vector {\tt naturalGroupProp}, (ii)    optional  arguments,  {\tt gammaMin} and {\tt gammaMax}, represent bounds for each element of the FLEXOR study proportions $\breve{\boldsymbol{\gamma}}$. In other words, $\Upsilon=\mathcal{S}_J \cap \bigl([\text{\tt gammaMin}, \text{\tt gammaMax}\bigr)]^J \times \{\boldsymbol{\theta}\}$ in the iterative steps  to estimate the FLEXOR pseudo-popuation, and (iii)   optional  argument  {\tt num.random}  indicates the number of random starting points for $(\boldsymbol{\gamma}, \boldsymbol{\theta})\in\Upsilon$. The  sample ESS maximized over  these {\tt num.random} independent replications identify  the estimated FLEXOR pseudo-population, for which the $N$ sample weights and  ESS comprise the function's output. 

 \bigskip

 For example, to calculate the FLEXOR weights, we load the  package and data, set a random seed to ensure reproducibility, and set the number of starting points for the iterative procedure before calling \texttt{balancing.weights()}:
 \begin{example}
R> library(WMAP)
R> data(demo)
R> set.seed(1)
R> num.random=25
R> output1 = balancing.weights(S, Z, X, method="FLEXOR", 
+                              naturalGroupProp=naturalGroupProp, num.random)

FLEXOR... estimate 10
FLEXOR... estimate 20 
\end{example}

The output \texttt{output1} is a result S3 list object of class ‘\texttt{balancing\_weights}’, which contains:
\begin{itemize}
    \item \texttt{wt.v}: the weights for each individual.
\begin{example} 
R> length(output1$wt.v)
[1] 450
\end{example}
    
    \item \texttt{percentESS}: the ESS of the FLEXOR weights.
    \begin{example}  
R> output1$percentESS
[1] 60.81412 
\end{example}
\end{itemize}

\subsubsection[Stage 2: causal.estimate()]{Stage 2: {\tt causal.estimate()}}  \label{sec:stage2}

For a prespecified pseudo-population,
 the function \texttt{causal.estimate()} first calculates the subject-specific normalized balancing weights and  sample ESS by a call to the  \texttt{balancing.weights()} function. Next, it estimates the means, standard deviations and medians of the counterfactual outcomes of the group $K$, in addition to the mean differences in the group. Finally, the function regenerates the bootstrap samples and estimates the same set of counterfactual features for the bootstrap samples. The input arguments are summarized in Table \ref{tab:stage2 inputs}. The workflow is illustrated for counterfactual means and SD in Algorithm \ref{alg:stage2}. The function returns a list of items summarized in Table \ref{tab:stage2 outputs}.

\begin{algorithm}[h]
\caption{Causal estimation of  counterfactual means and SDs}  
\label{alg:stage2}
\SetAlgoLined
\DontPrintSemicolon
{\textbf{Inputs$_2$:} {\tt S}, {\tt Z}, {\tt X}, {\tt Y}, {\tt B}, {\tt method}, 
{\tt seed}, {\tt parallel}, {\tt n\_cores}, {\tt outcome\_model}, 
{\tt crossfit\_folds}, {\tt naturalGroupProp}, {\tt num.random}, 
{\tt gammaMin}, {\tt gammaMax} 

\Comment{See Table \ref{tab:stage2 inputs} for input argument details}}   

    \SetKwFunction{stagetwo}{Causal.Estimate}
    \SetKwProg{Fn}{Function}{:}{}
    \Fn{\stagetwo{\textbf{Inputs$_2$}}}{

        \textbf{Inputs$_1 \gets$}   {\tt S}, {\tt Z},   {\tt X},  {\tt method},  {\tt seed},  {\tt naturalGroupProp},  {\tt num.random},  {\tt gammaMin},  {\tt gammaMax} 

        {\tt wt.v, percentESS} $\gets$ BALANCING.WEIGHTS(\textbf{Inputs$_1$}) \Comment{see Algorithm \ref{alg:stage1}}
        
        $\rho_1,\ldots,\rho_N \equiv$ {\tt wt.v}
         \Comment{empirically normalized weights}
        
         \If{{\tt outcome\_model = TRUE}}{
Replace {\tt Y} by fitted or cross-fitted outcome-model predictions
}

        \For{$z=1,\ldots,K$} {
        $\hat{\lambda}_{z} \gets \frac{\sum_{i=1}^N \rho_i\,Y_i\,\mathcal{I}(Z_i=z)}{\sum_{i=1}^N \rho_i\,\mathcal{I}(Z_i=z)}$ \Comment{estimated  $z$th group's counterfactual mean}
        
        $\hat{\sigma}_{z} \gets \biggl(\frac{\sum_{i=1}^N \rho_i\,Y_i^2\,\mathcal{I}(Z_i=z)}{\sum_{i=1}^N \rho_i\,\mathcal{I}(Z_i=z)} - \hat{\lambda}_{z}^2\biggr)^{1/2}$ \Comment{estimated  $z$th group's counterfactual SD}
        } %

    \vspace{12 pt}
    
    \For{$b=1,\ldots,B$} {
    Draw bootstrap sample {\tt S$_b$}, {\tt Z$_b$}, {\tt X$_b$}, {\tt Y$_b$}

    \textbf{Inputs$_{1b} \gets$}   {\tt S$_b$}, {\tt Z$_b$},   {\tt X$_b$},  {\tt method},  {\tt seed},  {\tt naturalGroupProp},  {\tt num.random},  {\tt gammaMin},  {\tt gammaMax} 
    
    {\tt wt\_b.v, percentESS\_b} $\gets$ BALANCING.WEIGHTS(\textbf{Inputs$_{1b}$}) \Comment{see Algorithm \ref{alg:stage1}}
    
    $\rho_{1b},\ldots,\rho_{Nb} \equiv$ {\tt wt\_b.v}
     \Comment{empirically normalized weights}
    
   \If{{\tt outcome\_model = TRUE}}{
Replace {\tt Y$_b$} by fitted or cross-fitted outcome-model predictions
}
    
    \For{$z=1,\ldots,K$} {
    $\hat{\lambda}_{zb} \gets \frac{\sum_{i=1}^N \rho_{ib}\,Y_{ib}\,\mathcal{I}(Z_{ib}=z)}{\sum_{i=1}^N \rho_{ib}\,\mathcal{I}(Z_{ib}=z)}$ \Comment{estimated  $z$th group's counterfactual mean}
    
    $\hat{\sigma}_{zb} \gets \biggl(\frac{\sum_{i=1}^N \rho_{ib}\,Y_{ib}^2\,\mathcal{I}(Z_{ib}=z)}{\sum_{i=1}^N \rho_{ib}\,\mathcal{I}(Z_{ib}=z)} - \hat{\lambda}_{zb}^2\biggr)^{1/2}$ \Comment{estimated  $z$th group's counterfactual SD}
    } %

    } %
    
       \textbf{return} {\tt percentESS}, estimates of means $\{\hat{\lambda}_{z}\}_{z=1}^K$, 
SDs $\{\hat{\sigma}_{z}\}_{z=1}^K$, bootstrap means 
$\{\hat{\lambda}_{zb}\}_{z=1}^K$, and bootstrap SDs 
$\{\hat{\sigma}_{zb}\}_{z=1}^K$. See Table \ref{tab:stage2 outputs} for all outputs 
of the actual implementation.
    }
\textbf{End Function}
\end{algorithm}

 Using the example  dataset included with the WMAP package,
we provide a step-by-step guide to the causal estimation  of different features of the group-specific counterfactual outcomes for the three weighting methods,  FLEXOR, integrative combined (IC), and integrative generalized overlap (IGO). 
As mentioned, the $K = 2$ groups of the  example dataset simulate the breast cancer subtypes, IDC and ILC, and the $L = 8$  outcomes corresponding to the mRNA expression levels of the targeted breast cancer genes, COL9A3, CXCL12, IGF1, ITGA11, IVL, LEF1,  PRB2,  and SMR3B. The goal is  unconfounded, covariate-balanced inference about  the group counterfactual  means, standard deviations, and medians, as well as counterfactual differences in group means and ratios of group standard deviations. 

 If necessary, load the necessary packages and data, set a random seed, and set the runtime parameters:
 \begin{example}
R> library(WMAP)
R> data(demo)
R> set.seed(1)
R> num.random=25
R> B=200
    \end{example}

 \begin{table}[!htbp]
 \setlength{\abovecaptionskip}{2pt}
 \centering
\caption{
 Input arguments for function \texttt{causal.estimate()}.}
 \label{tab:stage2 inputs}
\begin{adjustbox}{max width=\textwidth}
\begin{tabular}{lll}
\\
  \toprule
  \textbf{Argument}
  &\textbf{Short description} &\textbf{Default} 
		\\

\midrule
{\tt S} & Vector of factor levels representing the $N$  study &-\\ &  memberships.  Takes values in $\{1,\ldots,J\}$ \\
{\tt Z} & Vector of factor levels representing the $N$  &-\\
& group memberships.  Takes values in $\{1,\ldots,K\}$ \\
{\tt X} & Covariate matrix of $N$ rows and $p$ columns&-  \\
{\tt Y} & Matrix of $L$ outcomes, dimension $N\times L$&-  \\
{\tt B} &Number of bootstrap samples for variance estimation&0  \\
{\tt method} & Pseudo-population, i.e., weighting method;   &-  \\
&Can  be {\tt "FLEXOR"}, {\tt "IC"}, or {\tt "IGO"}\\
{\tt seed} & Seed for random number generation.   &{\tt NULL}  \\
{\tt parallel} & Logical; if TRUE, use parallel processing for   &{\tt FALSE}  \\
& bootstrap. \\
{\tt n\_cores} & Number of cores to use for parallel bootstrap.   &{\tt availableCores() - 1}  \\
& Only used if parallel = TRUE. \\
{\tt outcome\_model} & Logical; if TRUE, fits a random forest on X, S, Z and  &{\tt FALSE}  \\
& substitutes predicted outcomes for observed Y. \\
{\tt crossfit\_folds} & Integer $\geq$ 2 for K-fold cross-fitting of the  &{\tt NULL}  \\
& outcome model, or NULL to fit on the full data. \\ 
& Only used when outcome\_model = TRUE. \\
\midrule
\multicolumn{3}{c}{Relevant only when {\tt method="FLEXOR"}; inputs  ignored otherwise}\\
\midrule
{\tt naturalGroupProp} & Relevant only for FLEXOR pseudo-populations: &{\tt NULL} \\
&fixed user-specific probability vector $\boldsymbol{\theta}$ \\
{\tt num.random} & Number of random starting points of $\boldsymbol{\gamma}$ &40\\
& in the iterative procedure\\
{\tt gammaMin} &Lower bound for each $\gamma_s$ in the iterative procedure&0.001\\
{\tt gammaMax} &Upper bound for each $\gamma_s$ in the iterative procedure &0.999\\
	 \bottomrule
\end{tabular}
\end{adjustbox}
\end{table}

Then, call \texttt{causal.estimate()} setting \texttt{method} equal to \texttt{"FLEXOR"}, \texttt{"IGO"}, or \texttt{"IC"}. For example, the following command applies the FLEXOR weighting method.
\begin{example}
R> output2.f = causal.estimate(S, Z, X, Y, B, method="FLEXOR", parallel = TRUE,
+                              naturalGroupProp=naturalGroupProp, num.random)
      
\end{example}

\begin{table}[!htbp]
 \setlength{\abovecaptionskip}{2pt}
 \centering
\caption{
 Output list components of   function \texttt{causal.estimate()}.}
 \label{tab:stage2 outputs}
\begin{adjustbox}{max width=\textwidth}
\begin{tabular}{lll}
\\
  \toprule
  \textbf{Position}&\textbf{Names} 
  &\textbf{Short description} 
		\\

\midrule
1 & {\tt percentESS} &Percentage sample ESS of pseudo-population\\
2 & {\tt moments.ar} &Array of dimension $3 \times K \times L$, containing \\
 &  &$\bullet$ estimated means, SDs, and medians (dimension 1)  \\
 &  &$\bullet$ for $K$ groups (dimension 2)  \\
 & &$\bullet$ and $L$ counterfactual outcomes (dimension 3)\\
3 & {\tt otherFeatures.v} &Estimated mean group differences for $L$ outcomes\\
4 & {\tt collatedMoments.ar} &Array of dimension $3 \times K \times L \times B$, containing \\
 &  &$\bullet$  {\tt moments.ar} of $b$th bootstrap sample (dimensions 1--3)  \\
 &  &$\bullet$ for $B$ bootstrap samples (dimension 4)  \\
5 & {\tt collatedOtherFeatures.mt} &Matrix of dimension $L \times B$ containing\\
&  &$\bullet$  {\tt otherFeatures.v}   of $b$th bootstrap sample (dimension 1)  \\
&  &$\bullet$ for $B$ bootstrap samples (dimension 2)  \\
6 & {\tt collatedESS} &A vector of length $B$, containing \\
&  &$\bullet$ percentage sample ESS for $B$ bootstrap samples\\
7 & {\tt method} & Pseudo-population method, i.e., weighting method. \\
8 & {\tt selected\_nodesize} & Selected {\tt min.node.size} for the propensity forest \\
9 & {\tt diagnostics} & List of propensity model diagnostics, containing \\
  &  & $\bullet$ {\tt frac\_truncated}: fraction of subjects with near-zero\\ 
  &  & pre-truncation propensity \\
  &  & $\bullet$ {\tt crossfit\_accuracy}: OOF classification accuracy of \\ 
  &  & propensity model \\
  &  & $\bullet$ {\tt max\_weight}: maximum balancing weight \\
  &  & $\bullet$ {\tt prob\_quantiles}: quantiles of estimated propensity scores \\
	 \bottomrule
\end{tabular}
\end{adjustbox}
\end{table}

The output \texttt{output2.f} is a result S3 list object of class `\texttt{causal\_estimates}', which contains: 
\begin{itemize}
    \item \texttt{percentESS}: the ESS of the FLEXOR weights.
    \begin{example}
R> output2.f$percentESS
[1] 60.81412
    \end{example}

    \item \texttt{moments.ar}: the means, standard deviations, and medians of the mRNA expression of the 8 genes in the $K=2$ groups. 

    \begin{example}
R> output2.f$moments.ar
, , Y 1

           group 1    group 2
mean   -0.03186459 0.03946188
sd      0.98804608 0.60478468
median -0.11178418 0.18770154

, , Y 2

          group 1   group 2
mean   0.04682348 0.5700945
sd     0.98403675 0.7629177
median 0.14169894 0.6292224

...

, , Y 8

          group 1    group 2
mean   -0.6256443  0.0740488
sd      0.7175654  1.0337526
median -0.8660274 -0.3983188
\end{example}

    \item \texttt{otherFeatures.v}: the mean differences of the 8 genes between the two groups.
    \begin{example}
R> output2.f$otherFeatures.v
        Y 1         Y 2         Y 3         Y 4  
 -0.07132646 -0.52327102 -0.81733383  0.06576640  
         Y 5         Y 6         Y 7         Y 8 
  0.37685321 -0.33253838  0.02402617 -0.69969314 
    \end{example}

    \item \texttt{collatedMoments.ar}: the \texttt{moments.ar} for each bootstrap.
    \item \texttt{collatedOtherFeatures.mt}: the mean differences of the 8 genes between the two groups (\texttt{otherFeatures.v}) for each bootstrap sample. 
\end{itemize}

Based on the bootstrap results, we can calculate  95\% confidence intervals, for example, for the mean differences of the eight genes:
\begin{example}
R> CI.f = round(t(apply(output2.f$collatedOtherFeatures.mt, 1, 
+                   function(x) quantile(x,probs = c(0.025,0.975)))),2)
R> CI.f
     2.5% 97.5%
Y 1 -0.23  0.18
Y 2 -0.77 -0.35
Y 3 -1.03 -0.65
Y 4 -0.23  0.33
Y 5  0.21  0.57
Y 6 -0.64 -0.09
Y 7 -0.07  0.13
Y 8 -1.14 -0.45
\end{example}

To calculate the 95\% confidence intervals of the mean, median, and standard deviation for the mRNA expression levels of the eight genes in the two groups, we implement the following:
\begin{example}
R> f.moments.ci = apply(output2.f$collatedMoments.ar, c(1, 2, 3), function(x) {
+   quantile(x, probs = c(0.025,0.975))
+ })
R> f.moments.ci    
, , group 1, Y 1

             mean        sd      median
2.5%  -0.15730300 0.9050495 -0.22415331
97.5%  0.08789869 1.1294220 -0.03097463

, , group 2, Y 1

            mean        sd     median
2.5%  -0.1774918 0.4754648 -0.2005192
97.5%  0.1799238 0.8007878  0.2929509

...

, , group 1, Y 8

            mean        sd     median
2.5%  -0.6893114 0.5581658 -0.8972262
97.5% -0.4991113 1.0634283 -0.8350422

, , group 2, Y 8

            mean        sd     median
2.5%  -0.1627001 0.8566888 -0.6474411
97.5%  0.5490779 1.2679088  0.6007919
\end{example}

To include the 95\% CI in the output, we define a function \texttt{write\_res}:  
\begin{example}
R> write_res = function(estimates, CI){
+   lower_bound <- CI[, 1]  # 2.5% confidence bound
+   upper_bound <- CI[, 2]  # 97.5% confidence bound
+ 
+   sapply(1:length(estimates), function(i) {
+     paste0(round(estimates[i],2), " (", round(lower_bound[i], 2), ",", 
+            round(upper_bound[i], 2), ")")
+   })
+ }
\end{example}
and then use \texttt{write\_res} to output results including point estimates and CIs separately for each comparison group:

\begin{example}
R> f.moments = list()
R> for(i in 1:8){
+   f.moments[[i]] = cbind(group1 = write_res(output2.f$moments.ar[,1,i],
+                                   t(f.moments.ci[,,1,i])), # group 1 mean sd median
+                          group2 = write_res(output2.f$moments.ar[,2,i],
+                                   t(f.moments.ci[,,2,i]))) # group 2 mean sd median
+ }
R> f.moments
[[1]]
     group1                group2             
[1,] "-0.03 (-0.16,0.09)"  "0.04 (-0.18,0.18)"
[2,] "0.99 (0.91,1.13)"    "0.6 (0.48,0.8)"   
[3,] "-0.11 (-0.22,-0.03)" "0.19 (-0.2,0.29)" 

[[2]]
     group1              group2            
[1,] "0.05 (-0.1,0.17)"  "0.57 (0.42,0.75)"
[2,] "0.98 (0.85,1.09)"  "0.76 (0.62,0.91)"
[3,] "0.14 (-0.06,0.27)" "0.63 (0.41,0.81)"

...

[[8]]
     group1               group2             
[1,] "-0.63 (-0.69,-0.5)" "0.07 (-0.16,0.55)"
[2,] "0.72 (0.56,1.06)"   "1.03 (0.86,1.27)" 
[3,] "-0.87 (-0.9,-0.84)" "-0.4 (-0.65,0.6)"

\end{example}

For the other implemented weighting methods, i.e., IGO and IC, we would  change the \texttt{method} argument in \texttt{causal.estimate()} to  \texttt{"IGO"} and \texttt{"IC"}, and  follow the same procedures as above to obtain estimates and confidence intervals. More specifically, we apply the following commands:
\begin{example}
R> output2.igo = causal.estimate(S, Z, X, Y, B, method="IGO", parallel = TRUE, num.random)
R> output2.ic = causal.estimate(S, Z, X, Y, B, method="IC", parallel = TRUE, num.random)
    
\end{example}

\subsubsection{Computational Features}  \label{sec:compfeatures}

The \texttt{causal.estimate()} function supports optional bootstrapping via the \texttt{B} parameter. When \texttt{B = 0} (the default), the function returns point estimates only. When \texttt{B > 0}, the function performs \texttt{B} bootstrap iterations. The resulting bootstrap distributions are stored in the returned object and used to construct confidence intervals.

The package supports parallel bootstrap processing via the \texttt{parallel} and \texttt{n\_cores} parameters. The implementation uses the \texttt{future} and \texttt{future.apply} packages with automatic backend selection: \texttt{future::multicore} (forking) on macOS and Linux, and \texttt{future::multisession} (separate R sessions) on Windows and within RStudio, where forking is not supported. Reproducibility across parallel runs is ensured via \texttt{future.seed = TRUE}. The number of cores defaults to \texttt{parallelly::availableCores() - 1}, or can be set explicitly via \texttt{n\_cores}.

The \texttt{outcome\_model} parameter implements the optional outcome-modeling extension described in \hyperref[S:stage2]{Stage 2}.  When \texttt{outcome\_model = TRUE}, a random forest is fit separately for each outcome using covariates \texttt{X}, study membership \texttt{S}, and group membership \texttt{Z} as predictors, and the resulting predictions are substituted for the observed outcomes in the weighted estimator (\ref{estimator}).  By default \texttt{crossfit\_folds = NULL}, the random forest is trained on the full dataset. K-fold cross-fitting is available via the \texttt{crossfit\_folds} parameter. When bootstrapping is enabled, the outcome model is refit for each bootstrap iteration.

Propensity score estimation uses joint \texttt{ranger} probability forest. The \texttt{min.node.size} hyperparameter is selected via a tuning grid of \{10, 25, 50, 100\} using 5-fold cross-fitting with stratified fold assignment by (S,Z) cell. The smallest candidate satisfying three weight-stability criteria (fewer than 5\% of subjects with own-cell propensity below 0.01, maximum-to-median weight ratio below 10, and ESS/N above 20\%) is selected, and out-of-fold predictions from that candidate serve as the final propensity score estimates. Scores are truncated at a floor of 0.005 and renormalized to sum to one. \sg{Before the propensity forest is fit, the covariate matrix is replaced by a truncated singular value decomposition (SVD) reconstruction computed from the unscaled covariates, retaining the leading components that account for 60\% of the cumulative singular-value mass. Because the reconstruction is linear in the covariates, it preserves the linear component of the propensity signal in X. Nonlinear structure may be weakened when it is inadequately represented by the retained components, which limits the current implementation.  For bootstrap replicates, the selected {\tt min.node.size} is reused without additional tuning, and the propensity scores are obtained by the same cross-fitting procedure used for the point estimate, with all bootstrap copies of an original observation assigned to the same fold.}

The \texttt{balancing.weights()} function automatically checks for potential issues and issues warnings when: (1) the effective sample size falls below 5\% of the total sample size; (2) the ratio of maximum weight to mean weight exceeds 20; (3) any study-group cell contains fewer than 5 observations; or (4) more than 5\% of subjects have an estimated propensity score in their own (S,Z) cell below 0.005 prior to truncation. These warnings are informative but non-blocking, allowing users to proceed with caution and make informed decisions about their analyses. The returned object also includes the selected \texttt{min.node.size} and a \texttt{\$diagnostics} list containing fraction of subjects with own-cell propensity below the truncation floor prior to stabilization, out-of-fold classification accuracy of the propensity model, quantiles of the subject-level propensity distribution, and the maximum weight prior to normalization.

\subsection{Discussion of results}

We used \texttt{causal.estimate()} to compute point estimates and 95\% confidence intervals, and we summarize the results in Table~\ref{tab1}. The analysis was conducted on a simulated dataset modeled after a multi-site TCGA breast cancer study. While the dataset is not based on actual patient data, it reflects key structural and biological features commonly found in real-world genomic studies. This setup allows us to assess the performance of the methods under controlled conditions.

In this simulated setting, all three methods (FLEXOR, IC, and IGO) consistently indicated no significant difference in the counterfactual mean expression of the \textit{COL9A3} gene between the \sg{IDC and ILC} groups. However, \sg{all three methods indicated significantly greater variability in expression for IDC, with FLEXOR yielding the narrowest confidence interval for the ratio of group standard deviations. This finding is consistent} with the known biological heterogeneity of this subtype in real datasets \citep{wang2024progression}. This kind of variability could be important for understanding differential treatment responses or disease progression.

The genes \textit{CXCL12} and \textit{IGF1} showed lower counterfactual mean expression in IDC compared to ILC. These findings align with the biological functions of these genes; \textit{CXCL12} is involved in cell migration and metastasis, while \textit{IGF1} plays a role in cell growth and survival. Although these patterns were observed in simulated data, they demonstrate how meta-causal analysis can help detect group-level trends in gene expression that may be relevant in real studies \citep{vanden2005comparative}.

For both \textit{CXCL12} and \textit{IGF1}, the variability in expression did not differ meaningfully between subtypes. This suggests that while the mean expression levels diverged, the underlying mechanisms driving variability may be similar. In actual biological systems, this could reflect regulation through shared pathways or constraints on gene expression stability \citep{samani2007igf1_breast_cancer}.

Although the results are based on simulated data, they highlight how the methods in the \texttt{WMAP} package can support causal interpretation of group differences in gene expression, including both average levels and variability. These types of analyses may have practical applications in identifying subtype-specific markers or informing personalized treatment strategies \citep{mccart2021genomic}.

In nearly all scenarios, FLEXOR produced narrower confidence intervals than both IC and IGO. This higher precision illustrates FLEXOR’s strength in practice, particularly when theoretical conditions do not fully hold. By emphasizing stability and practical feasibility in weight estimation, FLEXOR provides a reliable approach to causal meta-analysis in genomic settings.

 \LTcapwidth=\textwidth

\begin{longtable}[t]{ccccc}
\caption{For three targeted genes, \textbf{COL9A3}, \textbf{CXCL12} and  \textbf{IGF1}, the estimates and 95\% bootstrap confidence levels (shown in parenthesis) of different population-level estimands of the potential outcomes of group 1 (IDC cancer subtype, denoted by superscript 1) and group 2 (ILC cancer subtype, denoted by superscript 2) with FLEXOR, IC, and IGO weights. 
An IC or IGO confidence interval is bolded if it is wider than the FLEXOR confidence interval. $\lambda$: group mean;  $\sigma$: group standard deviation; $M$: group median.}
\label{tab1}
\\
\toprule
\hline
\multicolumn{4}{c}{\textbf{COL9A3} } \\ \hline
\textbf{Estimand} & \textbf{FLEXOR} & \textbf{IC} & \textbf{IGO} \\\hline
$\lambda^{(1)}$ & $-0.03 (-0.16,0.09)$ & $0.12 \boldsymbol{(-0.13,0.33)}$ & $0.05 \boldsymbol{(-0.13,0.35)}$ \\
$\lambda^{(2)}$ & $0.04 (-0.18,0.18)$ & $0.06 \boldsymbol{(-0.29,0.18)}$ & $0.01 \boldsymbol{(-0.28,0.17)}$ \\ \hline
$\sigma^{(1)}$ & $0.99 (0.91,1.13)$ & $1.13 \boldsymbol{(0.91,1.28)}$ & $1.06 \boldsymbol{(0.82,1.28)}$ \\
$\sigma^{(2)}$ & $0.6 (0.48,0.8)$ & $0.71 \boldsymbol{(0.55,0.9)}$ & $0.72 \boldsymbol{(0.55,0.92)}$ \\ \hline
$M^{(1)}$ & $-0.11 (-0.22,-0.03)$ & $-0.04 \boldsymbol{(-0.17,0.06)}$ & $-0.06 \boldsymbol{(-0.23,0.15)}$ \\
$M^{(2)}$ & $0.19 (-0.2,0.29)$ & $0.13 \boldsymbol{(-0.34,0.3)}$ & $0.06 \boldsymbol{(-0.31,0.29)}$ \\ \hline
$\lambda^{(1)} - \lambda^{(2)}$ & $-0.07 (-0.23, 0.18)$ & $0.06 \boldsymbol{(-0.19, 0.41)}$ & $0.03 \boldsymbol{(-0.21, 0.45)}$ \\ 
${\sigma^{(1)}}/{\sigma^{(2)}}$ & $1.63 (1.25,2.13)$ & $1.59 \boldsymbol{(1.13,2.13)}$ & $1.46 \boldsymbol{(1.08,2.03)}$ \\ \hline
\multicolumn{4}{c}{\textbf{CXCL12}} \\ \hline
{\textbf{Estimand}} & {\textbf{FLEXOR}} & {\textbf{IC}} & {\textbf{IGO}} \\\hline
$\lambda^{(1)}$ & $0.05 (-0.1,0.17)$ & $-0.11 \boldsymbol{(-0.4,0.1)}$ & $-0.1 \boldsymbol{(-0.37,0.12)}$ \\ 
$\lambda^{(2)}$ & $0.57 (0.42,0.75)$ & $0.6 (0.45,0.77)$ & $0.65 (0.47,0.79)$ \\  \hline
$\sigma^{(1)}$ & $0.98 (0.85,1.09)$ & $1.07 \boldsymbol{(0.87,1.26)}$ & $1.03 \boldsymbol{(0.84,1.25)}$ \\ 
$\sigma^{(2)}$ & $0.76 (0.62,0.91)$ & $0.76 \boldsymbol{(0.63,0.94)}$ & $0.77 (0.63,0.88)$ \\  \hline
$M^{(1)}$ & $0.14 (-0.06,0.27)$ & $0.04 (-0.13,0.18)$ & $0.02 (-0.12,0.2)$ \\ 
$M^{(2)}$ & $0.63 (0.41,0.81)$ & $0.67 \boldsymbol{(0.38,0.84)}$ & $0.73 \boldsymbol{(0.38,0.84)}$ \\  \hline
$\lambda^{(1)} - \lambda^{(2)}$ & $-0.52 (-0.77, -0.35)$ & $-0.71 \boldsymbol{(-1.03, -0.45)}$ & $-0.74 \boldsymbol{(-1.00, -0.42)}$ \\
${\sigma^{(1)}}/{\sigma^{(2)}}$ & $1.29 (1.06,1.59)$ & $1.41 \boldsymbol{(1.06,1.75)}$ & $1.33 \boldsymbol{(1.04,1.76)}$ \\ \hline
\multicolumn{4}{c}{\textbf{IGF1} } \\ \hline
{\textbf{Estimand}} & {\textbf{FLEXOR}} & {\textbf{IC}} & {\textbf{IGO}} \\\hline
$\lambda^{(1)}$ & $0.13 (-0.05,0.23)$ & $0 \boldsymbol{(-0.17,0.16)}$ & $-0.01 \boldsymbol{(-0.23,0.21)}$ \\
$\lambda^{(2)}$ & $0.95 (0.8,1.08)$ & $0.85 \boldsymbol{(0.69,1.03)}$ & $0.89 \boldsymbol{(0.66,1.05)}$ \\ \hline
$\sigma^{(1)}$ & $0.97 (0.88,1.07)$ & $1 (0.9,1.06)$ & $0.99 (0.9,1.07)$ \\
$\sigma^{(2)}$ & $0.62 (0.47,0.78)$ & $0.7 (0.58,0.82)$ & $0.73 (0.57,0.83)$ \\ \hline
$M^{(1)}$ & $0.17 (0.01,0.32)$ & $0.05 \boldsymbol{(-0.09,0.23)}$ & $0.02 \boldsymbol{(-0.16,0.31)}$ \\
$M^{(2)}$ & $0.93 (0.83,1.08)$ & $0.91 \boldsymbol{(0.78,1.07)}$ & $0.94 \boldsymbol{(0.77,1.07)}$ \\ \hline
$\lambda^{(1)} - \lambda^{(2)}$ & $-0.82 (-1.03, -0.65)$ & $-0.85 \boldsymbol{(-1.07, -0.64)}$ & $-0.89 \boldsymbol{(-1.17, -0.60)}$ \\ 
${\sigma^{(1)}}/{\sigma^{(2)}}$ & $1.57 (1.24,2.1)$ & $1.43 (1.15,1.76)$ & $1.36 (1.13,1.76)$ \\ \hline
\bottomrule
\end{longtable}

\bigskip

\section{Simulations} \label{sim}

We used the WMAP package to generate simulated datasets for evaluating various weighting strategies used in inferring population-level characteristics across two subject groups. Specifically, we analyzed these datasets using the   \texttt{causal.estimate()} function. We briefly describe the generation strategy below and present the results. Following \cite{guha2024causal} and in alignment with the motivating TCGA breast cancer studies, we simulated $R = 250$ independent datasets, each containing $J = 7$ observational studies, $K = 2$ groups, and $L = 1$ (i.e., univariate) outcome for $\tilde{N} = 500$ subjects. The covariate vectors were sampled with replacement from the \texttt{demo} dataset included in the WMAP package, which mimics the structure of the $N = 450$ TCGA breast cancer dataset used in the motivating example. 
To demonstrate the practical application of the WMAP package, we present results from a representative simulation scenario designed to reflect key features of real-world retrospective studies with  confounding.
 This scenario effectively demonstrates the comparative performance of different weighting strategies implemented in the package. 
For the reported results, we used the procedure in Section~\hyperref[sec:methodology]{2} to meta-analyze the seven studies within each artificial dataset.

As an initial step for all $250$ artificial datasets, we conducted k-means clustering on the covariates, $\mathbf{X}_1,\ldots,\mathbf{X}_N$, from the WMAP \texttt{demo} dataset to identify lower-dimensional structure, grouping them into $Q=12$ clusters with centers $\mathbf{q}_1,\ldots,\mathbf{q}_Q \in \mathcal{R}^p$ and allocated covariate counts $m_1,\ldots,m_Q$. For each artificial dataset $r=1,\ldots,250$, containing $\tilde{N}$ patients, we then generated the data as follows:
 
\begin{enumerate}

\item \textbf{Natural population} \quad  For the $r$th artificial dataset, and using the Dirichlet distribution on simplex $\mathcal{S}_{Q}$, generate the relative masses of the $Q$ clusters, $\boldsymbol{\pi}^{(r)}=(\pi_1^{(r)},\ldots,\pi_Q^{(r)}) \sim \mathcal{D}_Q(\mathbf{1}_Q)$, with $\mathbf{1}_{Q}$ denoting the vector of $Q$ ones. Fix the patient population size of the large natural population as $N_0=10^6$, and 
sample their cluster memberships from a mixture distribution on the first $Q$ natural numbers: $c_{ir}^{(0)} \stackrel{\text{i.i.d.}}\sim \sum_{u=1}^Q \pi_u^{(r)}\zeta_{u}$, where $\zeta_{u}$ denotes a point mass at $u$. Select covariate $\mathbf{x}_{ir}^{(0)}$ uniformly from the $m_{c_{ir}^{(0)}}$ TCGA covariates assigned to the $c_{ir}^{(0)}$th k-means cluster (as above).

Generate the  group proportions in the natural population by drawing $\boldsymbol{\theta}^{(r)} \sim \mathcal{D}_K(\mathbf{1}_K)$, for $K=2$ groups. Define the group-covariate relationships: $\delta^{(r)}{z}(\mathbf{x}) \propto 1$ if $z=1$ and $\delta^{(r)}{z}(\mathbf{x}) \propto \exp\bigl(\omega_0^{(r)} + \omega_1^{(r)}\sum_{t=1}^p x_t/\frac{1}{N_0}\sum_{i=1}^{N_0}\sum_{t=1}^p x_{irt}^{(0)}\bigr)$ if $z=2$. Here, we set $\omega_1^{(r)} = 1$ and choose $\omega_0^{(r)}$ such that the population average of $\delta^{(r)}_z(\mathbf{x}_{ir}^{(0)})$ equals $\theta_z^{(r)}$.

\smallskip 

 \item \textbf{Covariates} \quad  
For each subject $i=1,\ldots,\tilde{N}$, select their covariate vector $\tilde{\mathbf{x}}_i^{(r)}=(\tilde{x}_{i1}^{(r)},\ldots,\tilde{x}_{ip}^{(r)})'$ by sampling with replacement from the $N=450$ \texttt{demo} dataset covariate vectors.

\item\label{sz combo} \textbf{Study and group memberships} \quad The study assignment $s_i^{(r)}$ and group label $z_i^{(r)}$ for each individual were generated as follows:
          \begin{enumerate}

                \item  \textit{MPS} \quad  
                Define the group-specific study propensities as follows:  
\[
\log \bigl(\delta_{S=s\mid Z=z}(\mathbf{x})/\delta_{S=1\mid Z=z}(\mathbf{x})\bigr) = sz\omega_1^{(r)}\sum_{t=1}^p\tilde{x}_{it}^{(r)}/\frac{1}{\tilde{N}}\sum_{i'=1}^{\tilde{N}}\sum_{t=1}^p\tilde{x}_{i't}^{(r)}
\]
for $s=2,\ldots,J$ and $z=1,2$. Assuming the group propensity scores  are the same as in the natural population, the marginal propensity score (MPS) is given by $\delta_{sz}(\mathbf{x})=\delta_{s\mid z}(\mathbf{x})\delta_{z}(\mathbf{x})$. For each patient $i=1,\ldots,\tilde{N}$, evaluate their probability vector $\boldsymbol{\delta}^{(r)}(\mathbf{x}_i) = \bigl(\delta_{11}^{(r)}(\mathbf{x}_i),\ldots,\delta_{JK}^{(r)}(\mathbf{x}_i)\bigr)$.
               
               \item \textit{Study-group memberships} \quad For each patient $i=1,\ldots,\tilde{N}$, generate $(s_i^{(r)},z_i^{(r)})$ from a categorical distribution with parameter $\boldsymbol{\delta}^{(r)}(\mathbf{x}_i)$.
            
\end{enumerate}

   \item \textbf{Subject-specific observed outcomes}\label{simulated response} \quad 
     Generate \[ Y_i^{(r)} \,\Big|\, Z_i = z_i^{(r)} \stackrel{\text{indep}}\sim \left( z_i^{(r)} \sum_{t=1}^p \tilde{x}_{it}^{(r)} + 50, \, \tau_r^2 \right), \] 
     where \( \tau_{r}^2 \) is selected to achieve an approximate \( R \)-squared of \( 0.9 \).

\end{enumerate}

\begin{figure}[h]
    \centering
    \begin{minipage}{0.33\textwidth}
        \includegraphics[width=\linewidth]{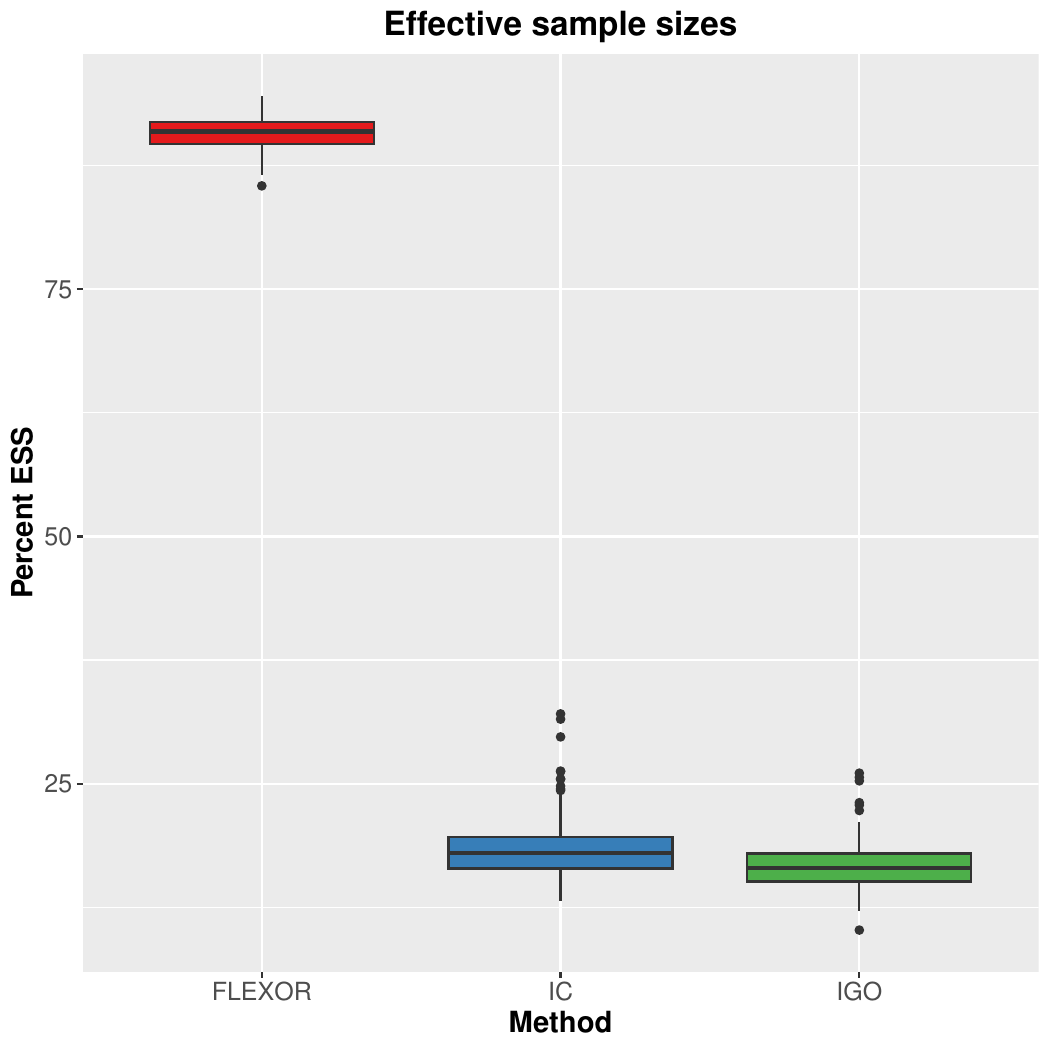}
    \end{minipage}%
    \begin{minipage}{0.33\textwidth}
        \includegraphics[width=\linewidth]{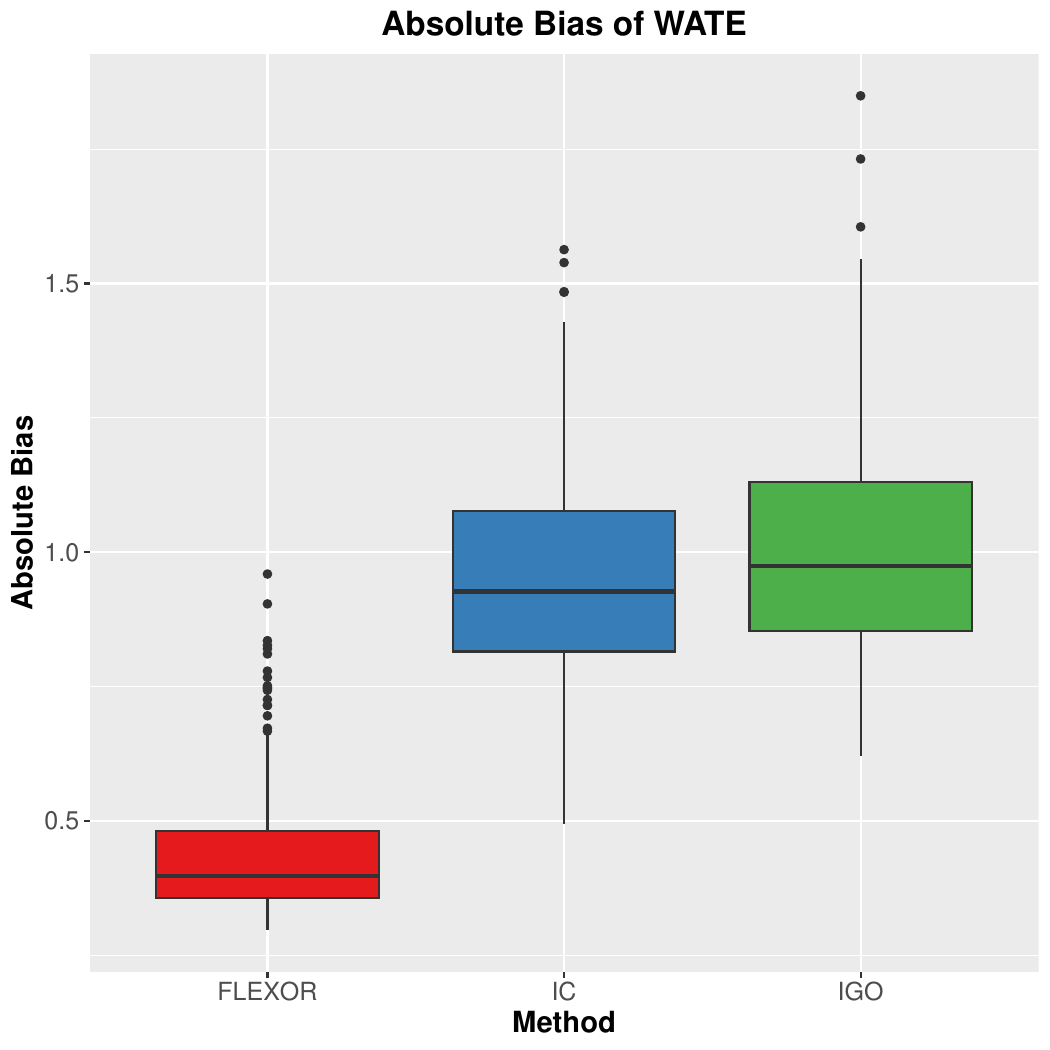}
    \end{minipage}
    \begin{minipage}{0.33\textwidth}
        \includegraphics[width=\linewidth]{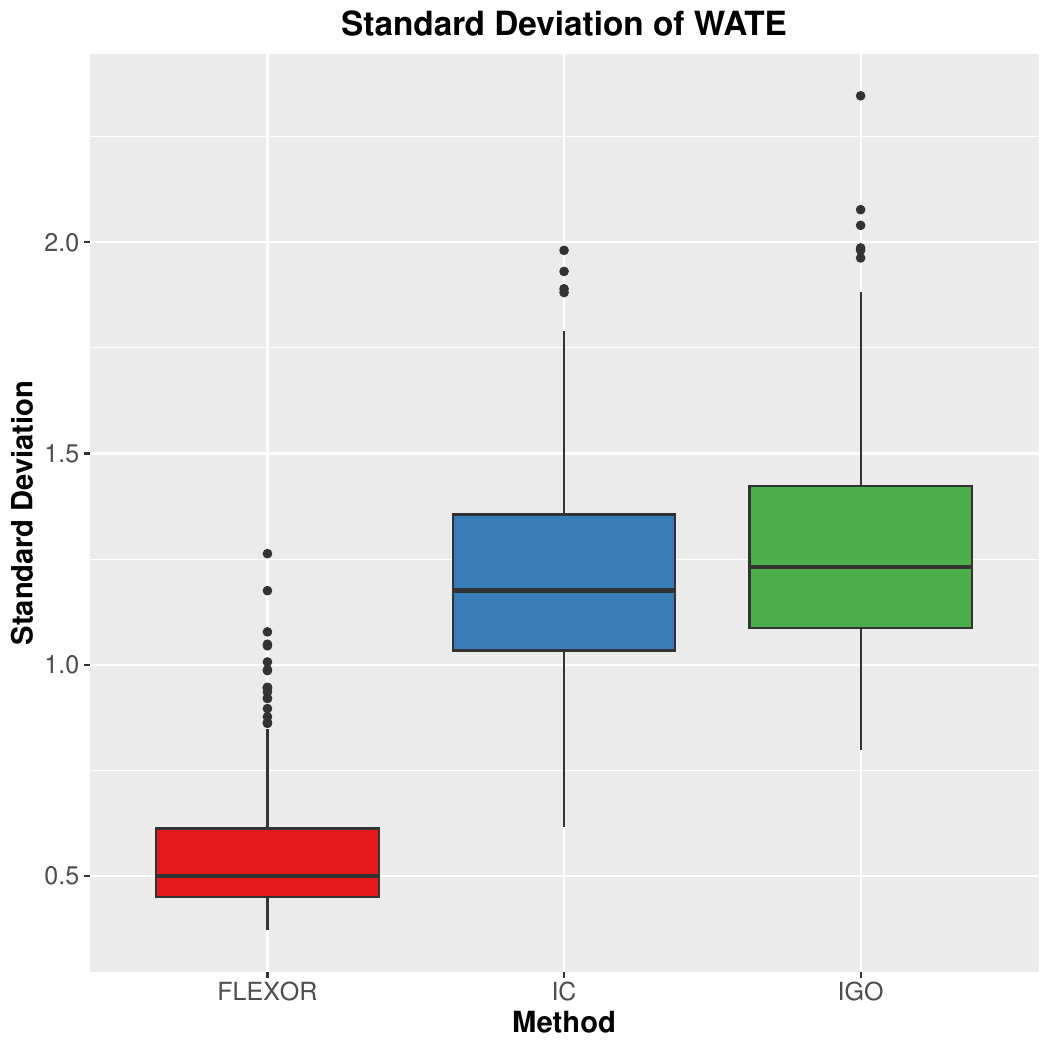}
    \end{minipage}
    \caption{Boxplots summarizing the results across 250 simulated datasets ($\tilde{N}=500$ subjects each): (i) percentage ESS, (ii) absolute bias, and (iii) standard deviation of the WATE (Weighted Average Treatment Effect) under three pseudo-populations.}
    \label{F:simulation}
\end{figure}

Next, we set aside knowledge of the simulation parameters and analyzed each artificial dataset using the procedure described in Section~\hyperref[sec:methodology]{2} for the IC, IGO, and FLEXOR pseudo-populations, as implemented in the WMAP package. 
 Define \textit{percent ESS} as the ESS scaled for 100 participants. For the 250 simulated datasets, the first panel of Figure \ref{F:simulation}  displays boxplots of the percent ESS for the FLEXOR, IGO, and IC pseudo-populations.  The IC and IGO pseudo-populations showed comparable ESS. The FLEXOR pseudo-population, however, consistently achieved substantially higher ESS across all datasets.

We employed the \hyperref[S:stage2]{Stage 2} strategy described in Section~\hyperref[sec:methodology]{2}, which is implemented in the WMAP package, to estimate the average treatment effect (ATE), defined as the difference in group-level counterfactual means, $\lambda^{(1)} - \lambda^{(2)}$. This was achieved by making weighted inferences under each method’s pseudo-population, where covariates were balanced to support causal interpretation. Because each estimator targets the ATE within its respective pseudo-population, we evaluated accuracy by comparing the estimated values to their corresponding true ATEs obtained via Monte Carlo simulation. The second and third panels of Figure~\ref{F:simulation} show the absolute biases and standard deviations of the FLEXOR, IGO, and IC estimators across 250 synthetic datasets. For each dataset and method, these performance metrics were computed using 200 independent bootstrap samples.

Overall, while the IGO and IC weighting strategies showed comparable performance in estimating the average treatment effect (ATE) across the simulated datasets, FLEXOR consistently outperformed both, as shown in Figure~\ref{F:simulation}.
 Additionally, it achieved lower absolute bias and standard deviation than the competing methods in all 250 artificial datasets. 
 Notably, although IGO weights are theoretically optimal for estimating the ATE under certain idealized conditions  for a specific target pseudo-population, such as homoscedastic outcomes and correct model specification \citep[see][for single studies]{li2019propensity}, these assumptions did not hold in our simulation setting. The strong empirical performance of FLEXOR in this context underscores its robustness and practical value, particularly when theoretical optimality conditions are violated. By placing greater emphasis on the stability and practical feasibility of the estimated balancing weights, FLEXOR offers a robust and reliable approach to ATE estimation in meta-analytical contexts, especially in scenarios where the assumptions underlying asymptotic efficiency may be less tenable.

\section{Conclusions and Future Developments}
\label{sec:conclusion}

Integrating multiple observational studies to make unconfounded causal or descriptive comparisons of group potential outcomes in large natural populations presents significant challenges, because of the complexities involved in data heterogeneity, selection bias, and the need for accurate balancing across datasets. Recently, \cite{guha2024causal} introduced a unified weighting framework designed to address these challenges by extending inverse probability weighting techniques for integrative analyses.
To translate this theoretical framework into practice, we have developed the
{R} package WMAP.
This software tool is specifically designed for the integrative analysis of user-specified datasets and implements three advanced weighting approaches, i.e., IC (Integrative Combined), IGO (Integrative Generalized Overlap), and FLEXOR (\sg{FLEXible, Optimized, and Realistic}). These methods enhance the capacity to estimate multiple propensity scores, compute balancing weights for subjects, evaluate effective sample sizes, and derive various estimands of counterfactual group outcomes. In addition, FLEXOR identifies efficient pseudo-populations satisfying user-specified population constraints. The package also includes functionality for calculating bootstrap variability estimates, which are essential for quantifying the uncertainty of the results.

In our illustrative application, we used WMAP to analyze simulated gene expression data that mimic differences between two major breast cancer subtypes: \sg{infiltrating} ductal carcinoma (IDC) and \sg{infiltrating} lobular carcinoma (ILC). Although based on synthetic data, the analysis yielded patterns consistent with known biological characteristics of these subtypes, demonstrating WMAP's potential to inform scientific insights in causal genomics. The package supports both observational studies and multi-arm randomized controlled trials (RCTs), especially when within-study treatment assignment mechanisms are known. Looking ahead, future updates will extend WMAP's functionality to accommodate hybrid study designs that integrate data from RCTs and retrospective cohorts. Current limitations, such as challenges in handling high-dimensional biomarker data, are being actively addressed. 
\sg{Another direction is extending the WMAP framework to remain robust under substantial violations of Assumption (I) in  Section~\hyperref[sec:methodology]{2}, which requires that, conditional on the covariates, study membership has no effect on the potential outcomes. This condition may fail in many integrative analyses. Relaxing it, for instance by allowing for unmeasured site-level heterogeneity in the potential outcomes, would broaden the applicability of the IC and IGO weighting approaches as well as FLEXOR.}
These enhancements aim to make WMAP an even more versatile and robust tool for integrative causal inference and descriptive analysis across diverse biomedical research settings.

\FloatBarrier

\section*{Declaration of generative AI use}
The authors used generative AI tools, including Claude (Anthropic) and ChatGPT (OpenAI), to assist with drafting and editing portions of the manuscript and with writing and debugging portions of the R package code. These tools were also used to transfer numerical results from output tables into the manuscript. All AI-assisted content was reviewed by the authors, who take full responsibility for the content of this article.
\newcommand{\noop}[1]{}

\end{document}